\documentclass[12pt]{article}
\usepackage{graphicx}
\usepackage{amssymb,amsmath,amsfonts,palatino,amsthm}
\usepackage{amssymb}
\usepackage{epstopdf}
\newcommand{\beq}{\begin{equation}}
\newcommand{\eeq}{\end{equation}}

\newcommand{\f}{\begin{equation}}
\newcommand{\ff}{\end{equation}}
\newcommand{\blankline}{\vskip .3cm}

\begin{document}

\title{The status of  cosmological natural selection }
\author{
Lee Smolin\thanks{Email address:
lsmolin@perimeterinstitute.ca}\\
\\
\\
Perimeter Institute for Theoretical Physics,\\
35 King Street North, Waterloo, Ontario N2J 2W9, Canada, and \\
Department of Physics, University of Waterloo,\\
Waterloo, Ontario N2L 3G1, Canada\\}
\date{\today}
\maketitle
\vfill
\begin{abstract}

The problem of making predictions from theories that have landscapes of possible low energy parameters is reviewed.  Conditions  for 
such a theory  to yield falsifiable predictions for doable experiments are given.  It is shown that the hypothesis of cosmological natural selection satisfies these conditions, thus showing that it is possible to continue to do physics on a landscape without invoking the anthropic principle.  In particular, this is true whether or not the ensemble of universes generated by black holes bouncing is a sub-ensemble of a larger ensemble that might be generated by a random process such as eternal inflation.  

A recent criticism of cosmological natural selection made by Vilenkin is discussed. It is shown to rely on assumptions about both the infrared and ultraviolet behavior of  quantum gravity that are very unlikely to be true.

\end{abstract}
\vfill
\newpage
\tableofcontents



\section{Introduction}

\begin{quotation}

{\it Nothing so dates an era as its conception of the future}

\blankline

		-Brian Eno

\end{quotation}

By a few years after the 1984 ``string revolution" it was clear that string theory would face an extreme version of the problem that philosophers call the problem of
underdetermination.  This was because it was quickly realized\cite{strominger}  that perturbative string theory would likely come in a vast, and perhaps infinite number of versions, generated by expansions around classical background geometries.   By analogy with population biology, the space of these theories  was 
called  the landscape \cite{LOTC}. Its existence may be called the {\it landscape problem.}  Put briefly, this is the question of whether {\it string theory or any theory based on a landscape can nonetheless generate robust and precise predictions that could be falsified by doable experiments,  based on our present knowledge of physics.}  

This issue compounded an already critical issue, which bedeviled attempts to explain the parameters of the standard models of particle physics and cosmology from a deeper theory.  This is the apparent fact that the parameters of the standard models of particle physics and cosmology appear to be chosen to be in a tiny subspace of the space of possible parameters which allows the existence
and copious production of long lived stars as well as a complex chemistry\cite{Rees,BT}.  Indeed, both long lived stars and stable nuclei owe there existence to highly improbable tunings of the parameters that give large hierarchies of dimensionless parameters.  We may call this the {\it special tuning problem}\footnote{It might be called the {\it anthropic observation} but this is misleading as the issues is the existence of carbon chemistry and stars, and not intelligent life, per se.}. 

A number of different approaches have been offered to the landscape problem and special tuning problem.  Some of them are based on one version or another of the anthropic principle\cite{Rees,BT}.  It is certainly true that in a weak form the anthropic principle makes logical sense: it is merely the combination of a very speculative hypothesis with some common sense advice. The speculative hypothesis is that the universe contains a vast number of regions where different laws operate, or where different parameters of those laws hold.  The common sense  advice is that selection effects have to be taken into account when computing probabilities in such an ensemble.  The problem with the anthropic principle in this  form is that it is nearly impossible to see how its use could lead to falsifiable predictions.  The reason is that there can be no independent confirmation of the speculative hypothesis about the existence of other universes.  Therefor one is free to entertain any assumptions about the ensemble of unobserved other universes that gives a desired result.   The poverty of this kind of approach has been demonstrated in detail 
in \cite{LOTC,AA,stevew,TTWP}. The arguments will not be repeated here, but it is worth mentioning  that every claim of a successful prediction based on the anthropic principle has been shown to be fallacious. One common fallacy is to append to an already complete argument a logically unnecessary statement about life and other universes.   The argument remains true but life had nothing to do with it.  Another common fallacy is to rely on assertions of typicality to convert extremely improbable outcomes to probable, where the notion of typicality is flexible enough that a large number of different outcomes can be obtained.  For more details see \cite{AA}.

Is there instead a possible explanation of the observed parameters of the standard model that solves the special tuning problem and is checkable by virtue of its making falsifiable predictions for doable experiments?  It is not difficult to discover that there are conditions under which this would be possible. These conditions are given in section 2.1 below.

Are there then examples of such possible explanations?  One of them was proposed in
\cite{evolve} and discussed in   \cite{CNS2,CNS3,LOTC}; it has come to be called {\it cosmological natural selection} (CNS).   This proposal is briefly reviwed in section 2.2.   The present status of the hypothesis with regard to predictions made in 1992, as well as challenges offered since, is reviewed in section 3.

Very recently A. Vilenkin has proposed an argument claiming to show that 
CNS is false\cite{Alex}.
In section 4.1 I present this argument and in section 4.2 explain why it fails.

\section{The motivation for cosmological natural selection}

\subsection{What is required for landscape theories to generate
falsifiable predictions?}

The original proposal of cosmological natural selection\cite{evolve} was motivated by early indications that string theory would have no unique physically preferred vacuum but instead have a vast number of equally physical solutions.  The notion of a landscape, $\cal L$  of theories was introduced by analogy to theoretical  biology, where the term ``fitness landscape" is in common use.  Reasoning in terms of a landscape of theories involves several structures:

\begin{itemize}

\item{}A space of fundamental theories or a parameter space of a single fundamental theory, denoted the fundamental landscape, 
$\cal L$.   On $\cal L$ is an ensemble of ``universes" 
${\cal E}_{\cal L}$ described by a probability distribution 
$\rho_{\cal L}$, 
which may depend on a global time parameter $t$.  $\cal L$ is
analogous to the space of genotypes in biology.  

\item{}There is a process that generates evolution in the ensemble,
leading to evolution of $\rho_{\cal L} (t)$ in $t$.  There may or may not be an equilibrium distribution which is constant in time $t$.  It is helpful to put a metric or topology on $\cal L$ so that two universes 
are closer if they have a better chance of evolving to one another.

\item{} We need also the space of parameters of the standard model
of particle physics, which we will call $\cal P$.  This is analogous to the
space of phenotypes in biology, as in that case it is relevant both because it is what we observe and because it and not the $\cal L$ may influence the evolution of the ensemble. 

\item{}There is a map ${\cal  I}: {\cal L} \rightarrow {\cal P}$.
The images of ${\cal E}_{\cal L}$ and ${\rho}_{\cal L}$ under
$\cal I$ give an ensemble and probability distribution 
${\cal E}_{\cal P}$ and ${\rho}_{\cal P}$ on $\cal P$.  This map may,
as in the case of biology, be highly non-trivial.  If, as is the case in biology and in the current models of the string landscape, 
$\cal L$ is a much larger space than $\cal P$ we may expect that 
$\cal I$ is onto a large region of $\cal P$ including the point
that labels the parameters of the standard model $p_{us}$ we
measure.  

\end{itemize}

In \cite{LOTC,AA} I argued that falsifiable predictions will be possible
in a theory based on a landscape if three conditions are satisfied:

\begin{enumerate}

\item{} The process $\cal M$ that generates the landscape leads to a probability distribution $\rho_{\cal P}$ which is {\it highly non-random}.  We assume as in standard uses of probability theory that 
our universe is a typical (that is randomly selected) member of this
ensemble. 

\item{}There will be observables $A_i$ whose  values in our universe are typical of those
of the ensemble specified by $\rho_{\cal P}$.   If so than the hypothesis $\cal M$ can be said in the context of this
scenario to explain the observed values of the $A_i$.  

\item{}There must be further properties, $B_i$ 
which have yet to be measured and which are not constrained by anthropic considerations but which are true in almost all universes
in the ensemble ${\cal E}_{\cal P}$ defined by $\rho_{\cal P}$.
In this case the theory predicts that our universe will have those
properties.  

In addition, a fourth condition is required if we want our theory to also solve the special tuning problem:

\item{}  To explain the special tuning observation the process
 that determines the evolution of the ensemble
${\cal E}_{\cal L}$ at a point $l \in {\cal L}$ must be highly sensitive to the low energy parameters at $p={\cal I}\cdot l $. Otherwise,  the process will not be able to generate universes preferentially in those regions of ${\cal P}$ that contain long lived stars and complex chemistry.   This is because those features are only found in universes where the low energy parameters fall into narrow ranges.  

\end{enumerate}

Note that these conditions all refer to the space of parameters of the standard model.  Assuming only that $\cal I$ is onto a region of
$\cal P$ containing $p_{us}$ we do not need information about the
fundamental landscape to test the theory.  But given information about
the map $\cal I$ some tests of this theory may imply tests of the fundamental theory.  

This is of course analogous to biology, in which many predictions could be made and tested without knowing the details, or even anything about, molecular genetics.  Indeed, the use of biology as an analogue is suggested by the fact that {\it evolutionary biology is the single case in science where a landscape problem and a special tuning problem have been solved successfully in a way that generated new falsifiable predictions.}  As argued in \cite{LOTC} this is probably not accidental, hence it makes good sense to try to apply the schema of evolutionary biology to cosmology.  

The claim made in \cite{LOTC,AA} is  {\it that no landscape theory will be able to either  produce  falsifiable predictions for doable observations or give a 
genuine explanation of the special tuning observation if it does not embody conditions 1-4. }  This is a prediction about theory;  I believe that so far, in spite of intensive investigation of alternative landscape theories over the last several years, it has stood up.  

\subsection{An existence proof for the argument}

Of course it might be that there are no theories that satisfy conditions 
1-4.  In this case we would have to conclude that  landscape theories are incapable of leading to falsifiable predictions.  Fortunately there is at least one example of a theory that satisfies 1-4, which is 
cosmological natural selection\cite{evolve,CNS2,CNS3,LOTC,AA}.  Its main importance is that it provides an existence proof that landscape theories can make falsifiable predictions.

The main hypothesis of cosmological natural selection are as follows:

\begin{itemize}

\item{}{\bf O}  The world consists of an ensemble ${\cal E}$ of universes, each of which  is characterized by a point $x \in {\cal L}$ and hence by 
a point $ p= {\cal I} x \in {\cal P}$.

\item{} {\bf I}.  Black hole singularities bounce and evolve to initial states of expanding universes\cite{wheeler}.  Hence there is a fitness function $f$ on ${\cal P}$ where $f(p)$ is equal to the average number of black holes produced by a universe initiated in such a bounce transition, which has parameters $p$.  

\item{}{\bf II.}   At each such creation event there is a small change
in $x$ leading to a small random change of $x$ in ${\cal P}$. Small here means compared to the step size under which $f(p)$ changes appreciably. 

\end{itemize}

Hence the ensemble ${\cal E}$ is updated by a discrete series of steps,
in each steps one new universe is created for each black hole in an existing universe. Under very mild assumptions for the fitness function one can show that after many steps the ensemble  converges to one in which almost every member is near a local extrema of
$f(p)$.  Hence the theory predicts that a randomly chosen member of the ensemble will have the following property, which we may call the {\it master prediction of CNS}

${\cal M}$: {\it Almost every small change in $p$ from its present value either leads $f(p)$ unchanged or leads to a decrease in 
$f(p)$.  Since our universe can be assumed to be randomly chosen we conclude that if the hypotheses made above are true, almost no change in the parameters of the standard model from the present values will increase the numbers of black holes produced. }

Note that $\cal M$ is a property of type {\bf B} as described in section 2.1 above.   It implies further type {\bf B} properties, and these, as we will review shortly,  become the falsiable predictions the theory makes.  

We may note that the theoretical evidence for the hypotheses of CNS have strengthened greatly since they were made in 1992.  There is, to put it mildly, a far greater consensus that string theory provides a landscape of theories\cite{KKLT}.  And there is much more theoretical evidence that both black hole and cosmological singularities bounce\cite{bojowald}.  

It is important to note also that the master prediction $\cal M$ refers to the observed low energy parameters and not to the fundamental landscape of theories. This allows it to be tested in spite of our ignorance about the details of the landscape and the map $\cal I$
between the landscapes of the fundamental theory and the low energy parameters. In our present state of knowledge this is an advantage, for the same reason that evolutionary biology could imply predictions that were successful in the absence of any knowledge of the molecular basis of genetics. 

Note also that $\cal M$ refers only to local changes around the observed values of the standard model parameters.  It is commonly asked why one cannot assert that the theory predicts that the population is peaked around a global maximum of $f(p)$. The reason is that this requires some detailed knowledge of the properties of the two landscapes and the map $\cal I$ which we do not have at the present time.  

Indeed, the whole point of CNS is to demonstrate that we can make genuine falsifiable predictions from landscape based theories in the absence of detailed knowledge of the fundamental landscape.   This was certainly the situation in 1992 when the theory was constructed.  Given more knowledge about the properties of the string theory landscape gained in recent and future investigations, more may become possible. But the claim remains that any such prediction requires the conditions specified above.  

Finally, note how the fact that the probability ensembles generated are highly non-random does work here to protect the theory from the kinds of ambiguities that plague applications of the weak anthropic principle.  In the AP case the theory generates an ensemble which is random on $\cal L$ and $\cal P$, within which a universe like ours is highly improbable.  One must then add supplementary assumptions such as the {\it principle of mediocrity} to pick our universe out of the tail of a distribution where it is very untypical.  The result is that the resulting claims tend to be highly dependent on the precise characterization of the ensemble in which our universe is expected to be typical. As demonstrated in real cases such as ``predictions" for the cosmological constant, one can get very different predictions depending on which parameters are allowed to vary in the choise of ensemble in which we are stipulated to be typical.  This does not happen in CNS, because our universe is assumed to be typical within the original ensemble generated by the theory there is no use for a further principle of mediocrity.  Hence, the master prediction
$\cal M$ depends on very little information about the landscape, but it is sufficient to yield precise falsifiable predictions, as we will now see\footnote{This is the answer to some comments at the end of Vilenkin's \cite{Alex}.  The full argument with references is in \cite{AA}.}.

\section{Successes of cosmological natural selection}

Cosmological natural selection would have done important work, even if it had been falsified, because its main motivation, initially, was to provide an illustration that theories can be formulated that satisfy the conditions given above for landscape theories that allow falsifiable predictions to be made.   But it is remarkable that, at least to date, it has been more successful than that, as it has survived falsification in spite of being vulnerable to it by ongoing observations.  

\subsection{Explanations}

CNS, if true, solves the special tuning problem mentioned 
above\cite{evolve,CNS2,CNS3,LOTC}.  It does 
so because the production of massive stars, required for copious production of astrophysical black holes turns out to require carbon chemistry. This is because the dominant cooling mechanism in the giant molecular clouds where massive stars appear to form is radiation from molecular vibrations of $CO$.  Furthermore, the giant molecular clouds require shielding from uv light from the massive stars which is provided by carbon dust and ice.   

This is to say that CNS explains why the universe is tuned so that there are stars and carbon chemistry.  This is the only explanation ever offered for this fact that is not anthropic-ie does not use the existence of life as a part of the explanation.  

CNS also explains a number of other coincidences.  One of these is that the Fermi constant is in the narrow range required for supernovas to work.

\subsection{Predictions}

Three predictions have been published about 
cosmological natural selection\cite{evolve,CNS2,CNS3,LOTC}.  To my knowledge all of these so far hold up.  

\begin{enumerate}

\item{} CNS implies a prediction that neutron stars are Kaon-condensate stars and that the upper mass limit for neutron stars is $M_{uml} \approx 1.6 M_{solar}$\cite{bethebrown}.   This comes about because the strange quark mass can be varied to raise and 
lower $M_{uml}$ without strongly affecting the processes that lead to massive star formation and supernovas\cite{evolve}. 

So far all the well measured neutron stars have masses between $1.3$ and $1.45$ solar masses. There is one dangerous case of a neutron star in which imprecise estimates give a range of masses which exceeds the prediction, but at less than one sigma\cite{estimates}.  

\item{} CNS implies that inflation, if true, must be governed by a single parameter, so that the inflaton coupling that controls $\delta \rho \over \rho $ also controls the number of efoldings $N$ \cite{evolve}.  When this is true one cannot just increase  $\delta \rho \over \rho $ to produce a large number of primordial black holes, because the result is a lower value of $N$ and hence an exponentially smaller  universe which produces many fewer black holes overall.  But in more complicated inflation models the parameters that control
 $\delta \rho \over \rho $ and $N$ are independent.  Hence, CNS predicts that the correct model of inflation must be the one in which these parameters are anti-correlated as in the standard single field inflation. 
 
 So far the predictions of single field, single parameter inflation hold up very well. 

\item{} Little early star formation.  The explanation of the special tuning observation given by  CNS could be wrong is if there are channels for massive star formation other than those presently observed,  which do not require carbon chemistry\cite{evolve}.  But if this were the case they might operate at high $z$ when the abundances of carbon and oxygen are much lower.   In that case it might be possible to observe many more supernovas at high $z$.  These so far have not been observed\cite{early}.

\end{enumerate}

It can be noted that these are the only predictions, falsifiable with current observations,  that have ever been made from a landscape theory.

\subsection{Prior attempted refutations}

 There have also been attempts at refutations(\cite{ellis}-\cite{rees}) that, on examination, did not succeed because they rested on too naive analyses of astrophysical processes.  These include variations of Newton's constant, of the baryon density and of the initial mass function.  These are discussed in \cite{LOTC,CNS3}.  

Another objection has been voiced in conversation but not, I believe, in print\cite{Linde-pc}.
This is that eternal inflation (EI) and CNS cannot be  both true because the mechanism of universe generation employed in EI would, if true, overwhelm the ensemble of universes created.   The reason for this expectation is the vast numbers of universes created with a very short time scale in eternal inflation.  

Before answering one should note that  eternal inflation alone is fated to employ the anthropic principle and some version of the principle of mediocrity if it is to solve the special  tuning problem.  This is because the mechanism of eternal inflation takes place at grand unified scales much higher in energy than the natural scales of nuclear and atomic physics, hence they are insensitive to whether the distribution of parameters is friendly to chemistry and long lived stars.  Thus, it is subject to the problems analyzed in \cite{AA} which show that it is unlikely to either solve the special tuning problem or be the basis of falsifiable predictions.  

But the assertion that CNS and EI are incompatible is simply false. First of all, both mechanisms of universe generation may in fact function in our universe.  The problem is then how to pick the structured sub-ensemble generated by CNS out of the much larger random ensemble generated by EI\footnote{As we just noted, one always has to define a sub-ensemble in EI because the ensemble EI generates is one in which our universe is very atypical.}.  But this is easy, one just has to ask the right questions.
In a world where CHS and EI both function a property a universe will have is $N$, the number of ancestor universes it has from black hole bounces.  One can consider the sub-ensemble in which $N$ is larger than some very large $N_0$. Let us call this sub-ensemble ${\cal E}_{\cal L}^{N_0}$.  We can posit that our universe is a typical member of it.  The reasoning of CNS can then be applied just to this sub-ensemble, and the 
master prediction $\cal M$ will be found to apply to it.   

But is the assumption that our universe is in the sub-ensemble ${\cal E}_{\cal L}^{N_0}$ independently checkable, and does its postulation weaken the emperical content of the theory?   The answer is that it does not because it is already assumed that the statistical reasoning used in CNS is applied to the ensemble created by many generations of bouncing black holes.  This is logically equivalent to postulating that our universe is in ${\cal E}_{\cal L}^{N_0}$.  Whether that ensemble is a sub-ensemble of a much larger ensemble generated by EI or any other random process plays no role in the argument.

\section{A reply to Vilenkin's argument}

Very recently Vilenkin proposed an argument which he claims falsifies
CNS \cite{Alex}.  I first introduce the argument, then explain why it is much weaker than may appear at first.  

\subsection{Vilenkin's argument}

The basics of the argument is very simple and is as follows:

\begin{enumerate}

\item{}Experimental evidence points to the existence of a small cosmological constant, $\Lambda_0 \approx 10^{-122} l_{Pl}^{-2}$.  

\item{}Assuming, first, that $\Lambda_0$ is unchanging throughout the lifetime of the universe, the universe will after a few tens of billions of years, and for ever after,  be described to a good approximation by vacuum eternal deSitter spacetime.  

\item{}Eternal deSitter spacetime has a finite temperature 
\f
T_0 =\frac{1}{2\pi R} 
\label{dStemp}
\ff
 where $R^{-1} = H = \sqrt{\frac{\Lambda}{3}}$
is the deSitter horizon scale.  

\item{}In thermal equilibrium there is a very small, but nonzero rate for
eternal vacuum deSitter to nucleate black holes by quantum fluctuations
(For some observers this will look like nucleation by thermal fluctuations due to the intrinsic deSitter temperature).  This has been computed by Ginsparg 
and Perry\cite{GP} and others\cite{others} using the Euclidean path integral to semiclassical order, and the rate to nucleate a  black hole of mass $M$ per volume and per time found to be
\f
\Gamma = \frac{1}{l_{Pl}^4}e^{-\frac{M}{T_0}}
\label{nucl1}
\ff
This is maximized by a Planck mass black hole, for which the rate is
\f
\Gamma|_{max} = \frac{1}{l_{Pl}^4}e^{-\frac{2\pi R}{t_{Pl}}}
\approx \frac{1}{l_{Pl}^4}e^{-10^{61}}
\label{nucl}
\ff

\item{}  This is a very tiny rate. But because deSitter space increases its volume exponentially there will be a time $t_N \approx R \ 10^{61}$ at which the first
nucleated black hole is expected to appear.  Moreover given the exponential expansion, not too long after that a great number of black holes appear.  These
will swamp any amount of black holes produced by ordinary astrophysical processes. 

\item{}Thus, as long as the universe is in the deSitter vacuum for a time of
at least, 
say, twice $t_N$, and as long as the calculation of Ginsparg and Perry is reliable.  The overwhelmingly dominant contribution to black hole production up till $2 t_N$ will be spontaneous nucleation of black holes
by the Ginsparg-Perry process.  However, we see from (\ref{dStemp})
and (\ref{nucl}) that the number of black holes produced by any finite
time after $t_N$ is strongly increased by making a small increase
in $\Lambda_0$.  Hence, {\it given that the above assumptions hold,
the present universe is far from optimised for the production of black holes.}

\item{}Given this, the argument will hold in any landscape theory in which there are
metastable deSitter vacua with lifetimes long compared to $t_N$.  As $t_N$ is less
than the Poincare recurrance time, $t_P \approx R 10^{120}$ it is expected that there will be vacua in the string theory landscape that satisfy this\cite{KKLT}. 

\end{enumerate}

\subsection{Critique of Vilenkin's argument}

I have put Vilenkin's argument in a somewhat stronger form than he did.
In particular, there is no reason to talk of infinite production over infinite time, it is sufficient for his argument that there is a finite time after which the 
Ginsparg-Perry process dominates the production of black holes.   Indeed, if we accept the assumptions of their argument, the conclusion follows.    

\subsubsection{Preliminaries}

There are a few preliminary remarks that should be made.  First, strictly speaking, what the master prediction $\cal M$ implies is only that most parameters cannot be varied to increase the number of black holes.  If there were only a single such parameter, and all the rest of them turned out to agree with $\cal M$ that would be a remarkable fact and one would be tempted to still take the theory seriously. After all, in biology, real organisms are not totally optimized, one can always find a few mutations that would increase their fitness.   But as we shall see it is not necessary to duck out of Vilenkin's argument in this way.

Second, it is important to emphasize that all that CNS predicts is the behavior of small changes in the low energy parameters, around our present values, $p_{us}$.  Vilenkin's argument appears to fulfill this, as he needs only small changes in $\Lambda$ to produce
exponentially large changes in the numbers of black holes produced at any finite time
$t >> t_N$.  This is indeed why the argument must be rephrased in terms of a finite time, because all infinite ensembles have the same cardinality.  

Third, while it appears at first that Vilenkin's argument is safe from Planck scale physics, where we expect it breaks down, this is not in fact the case. The ensemble (\ref{nucl1}) that is produced is dominated by Planck mass black holes, hence it is dominated by processes which are sensitive to the ultraviolet completion of general relativity.   Vilenkin may want then to insist only on the minimal production of black holes, given for the lowest
possible mass black holes, when  $R_{Sch}=R$. The problem is then that $t_N$ is exponentially longer, and  is in fact equal to the Poincare recurrence time, 
$t_P \approx R \ 10^{120}$ at which no 
member of the string theory landscape is expected to be stable\cite{KKLT}.   This is to say that the reliability of his argument is constrained by the fact that we expect new physics for both ends of his spectrum,
for both $M=M_{Pl}$ and $M= R$.  

\subsubsection{The problem of extrapolating current laws over vast scales}

It is important to note that  Vilenkin's argument is not a falsification in the sense of the above discussion: a falsification is an instance in which a theory makes a prediction for an experiment which, when done, returns an outcome that contradicts that prediction.  Instead what we have here is a purely theoretical argument. The above arguments involve black holes that we have good observational evidence for.   The Vilenkin argument asks us to believe in a new mechanism of formation of black holes that-even under the assumptions of the argument-will not result in even a single black hole being produced till a span of time has gone on which is vastly longer than the present age of the universe. Indeed, the ratio of the time scale to create a single one of Vilenkin's black holes to the present age of the universe is the same as the Planck time to the latter.

We know that the laws that govern the present cosmological scale are very different from the laws that dominate on Planck scales. But even now there are great uncertainties about the laws that govern both the cosmological scale and any scale shorter than $10^{17} cm$.  Vilenkin's argument presumes that we can apply laws that govern presently experimentally accessible time scales to scales $10^{60}$ times bigger.  How likely is it that such reasoning is reliable?

What is required to be stable over such vast changes of scale is the least understood of the low energy parameters, which is the cosmological constant. If we had to pick out the single parameter about which we expect revolutionary insights, it is this one.  We may then anticipate that new insights concerning the cosmological constant and the dark energy will be necessary as physics progresses and these are likely to affect the assumptions of Vilenkin's argument.  

The actual fact of the matter is that more than 30 years after the standard model of particle physics was written down, we have no confident prediction for post-standard model physics that may be operating at scales just a single order of magnitude higher in energy than we have already probed.   What accounts for the dynamical symmetry breaking in the electroweak interactions and what determines the physical scale where it happens?  We really don't know, it could be supersymmetry, technicolor, preon models, dynamical symmetry breaking driven by top-antitop condensation, just to mention a few possibilities.  All we know is that presently none of these work very well, all of them, including supersymmetry and technicolor, which are the most beautiful options, are pushed into corners of their parameter spaces.  

Certainly, we should not be surprised as it has hardly ever been the case that we could extrapolate current laws more than one or two orders of magnitude before encountering unexpected phenomena.   But if we cannot reliably extrapolate current knowledge a single order of magnitude how likely is it we can extrapolate sixty orders of magnitude?

Vilenkin might reply that point $7$ above requires only that there be in the string landscape somewhere theories that satisfy the assumptions given.  But this is in fact too strong, for the reason that the master prediction $\cal M$ above is restricted to small changes in the 
parameter space $\cal P$ of low energy parameters.  So even if there are points in the string theory landscape $\cal  L$ that resemble the deSitter vacua for times larger
than $t_N$ we must show more to have an argument relevant for the truth of CNS (because nothing in CNS precludes there being universes far from our present parameters in 
$\cal P$ which produce many more black holes than our universe.)  We must show that  
{\it ${\cal I}^{-1} $  maps a small neighborhood of the low energy parameter space $\cal P$ around $p_{us}$ to a region in $\cal L$ where deSitter vacua that are stable on times
longer than $t_D$ are probable.} 

This means that CNS makes a prediction, which is that the statement just put in italics is not true of the real landscape.  

\subsubsection{Evidence of new phenomena already at the present Hubble scale}

Another vulnerability in Vilenkin's argument is that it  requires that the framework of a $\Lambda CDM$ universe should be reliable  for time scales up to $t_N$.  This means  it certainly must hold up to all presently observable scales. 

In fact, we already have evidence that the $\Lambda CDM$ model may be breaking down at the present Hubble scale.   While the predictions of inflationary cosmology, particularly $\Omega =1$ and a slightly red spectrum of fluctuations are confirmed at scales below the present Hubble scale, there is a list of anomalies that contradict the basic prediction of inflation that near scale invariance, homogeneity and isotropy govern much larger scales. These include not only the apparent lack of power on scales greater than $60$ degrees\cite{lackpower}, but also the existence of large anisotropies for modes between $l=2$ and $l=5$, the 
so called ``axis of evil"\cite{axisofevil}, and the apparent fact that up to modes of $l=40$ half the sky is hotter than the other half\cite{brighthalf}.

These may all be shown to be problems with the data analysis or instrument (although they are there also in the COBE data), but attempts to resolve them so far have failed.  At the very least, it can be said that the data points more to new,unexpected phenomena at scales of our present $\Lambda$ than it does to the continued extrapolation of current laws to larger distance scales.  Yet, to accept Vilenkin's argument as likely requires us to continue present laws another $60$ orders of magnitude beyond that scale.   

\subsubsection{The dependence of Vilenkin's argument to the  choice of the infrared completion of general relativity}

There is indeed good theoretical evidence that our understanding of gravity and cosmology needs an infrared completion in the shape of modifications of general relativity and/or quantum theory.  These include first of all the need to both solve the cosmological constant problem and account for the observed accelerating expansion.  There are several intriguing suggestions for such modifications which are under investigation\cite{modifications,nonlocal}.   These include quintessence, a ghost condensate, new terms in the Einstein equations, a role for  higher dimensions, or of new non-local effects\cite{nonlocal}.  All of them can account for the observed expansion as well as a cosmological constant. 

Indeed, the option of just adding a cosmological constant is also a modification of general relativity motivated by the data.  It is the simplest such extension, but it is far from the only plausible one. From a theoretical point of view the cosmological constant is indeed the least likely of the possible extensions which account for the data because it offers absolutely nothing in the way of either accounting for the new phenomena or explaining the observed value of $\Lambda_0$, whereas several of the alternatives do both.  Moreover, by doing so the alternatives have the great advantage of predicting new phenomena which may be observed in future data. 

But on any of the alternatives the calculations in Vilenkin's argument would come out differently because they predict an evolution for the universe on scales much larger
than $\Lambda^{-\frac{1}{2}}$ very different from that given by general relativity.  Some of these are quite radical, for example, a dissolution of local physics by a growing dominance of non-local effects\cite{nonlocal}.  But others are quite conventional, for example in quintesssence models the cosmological constant may simply decay in time to zero, well before the era of black hole  nucleation.  

\subsubsection{Problems with the use of the Euclidean path integral in quantum gravity}

Even within the framework of general relativity and quantum field theory there are serious worries with the assumptions of Vilenkin's argument.  Two of these come from his use of the Euclidean path integral to compute the nucleation rate for black holes in deSitter spacetime.   

The use of the Euclidean path integral to compute the thermal Greens functions in any quantum system, including general relativity, rests on a prior assumption, which is that the system in question has come to thermal equilibrium.   But it is well known that the gravitational degrees of freedom never reach thermal equilibrium\cite{noequil}.   It is not difficult to show that once produced, the mean free time for a graviton, $\tau (t)$  to have interacted with anything, measured at any cosmological time $t$ always satisfies:  
\f
\tau (t) > t
\ff
The universe either recollapses before the graviton interacts or the universe dilutes so fast it never interacts.  Another way to argue this is to show that any region of the universe dense enough to contain gravitational radiation long enough for it to interact and come to equilibrium will have collapsed to a black hole before those interactions take place.   These calculations tell us that it is incorrect to assume that the gravitational degrees of freedom come to equilibrium in any context where their dynamics is governed by general relativity.  

But the use of the Euclidean path integral with periodic imaginary time $t_I$ to compute thermal Green's functions  rests on the $KMS$ condition which assumes that the system is in thermal equilibrium\cite{KMS}.  There is no justification for using it to study a system that is not in thermal equilibrium\footnote{This is also seen concretely in \cite{laura}}.

This is not the only problem with the use of the Euclidean path integral in quantum gravity.  Its use, as proposed by Gibbons, Hawking, Hartle and others in the 1970s rested on the assumption that one could compute amplitudes in Lorentzian quantum gravity by doing the Euclidean path integral and then continuing the resulting amplitudes back to Lorentzian time.  But this assumes that the Euclidean path integral itself makes sense. In fact, we know now that the Euclidean path integral does not define a sensible theory beyond the semiclassical approximation. This is because  Euclidean quantum gravity above 2 dimensions has no critical phenomena that could be used to define the continuum limit.  This is shown using  dynamical triangulation methods to regularize the path integral\cite{dyntri}.  The results show that  there is no way to tune the parameters of the Euclidean path integral above $d=2$ while taking the continuum limit in such a way that a large scale classical spacetime geometry results.   

On the other hand, there is recent evidence that the Lorentzian path integral for general relativity can be defined directly in such a way that a continuum limit exists, and such that the infrared behavior of the theory so defined is consistent with the existence of a large scale classical spacetime geometry\cite{renate2,renate4}.   This work explicitly shows that the Euclidean path integral does not define a theory in the same universality class as the Lorentzian path integral.   This is by now welll understood in $d=2$\cite{renate2} and there is good numerical evidence for this conclusion as well in $d=4$\cite{renate4}.  

This is an example of a general phenomena in quantum gravity which is uv-ir mixing. Considerations necessary to define a theory in the ultraviolet do effect the infrared limit of gravity theories.  Another example of this is the requirement that the local gauge group, whether Lorentz or deSitter,  in a quantum gravity theory is quantum deformed\cite{qdef,positive} when the cosmological constant is non-zero.  This can be seen in $d=3,4$ from several different calculations. It is required also to make the Hilbert space defined on the horizon of deSitter spacetime finite dimensional, so as to be consistent with the holographic hypothesis\cite{positive}.  This in turn affects the low energy limit of the theory for arbitrarily small $\Lambda$.  For example a consequence in $d=3$ and quite possibly $d=4$ is that the contraction of the low energy symmetry as $\Lambda \rightarrow 0$ is not the Poincare algebra but a quantum deformed algebra: 
$\kappa-$Poincare\cite{DSR2+1,me-semi}.  But this effect is completely absent in the semiclassical approximation Vilenkin's argument refers to, hence that approximation misses phenomena dominant in the low energy limit of the correct theory.   

Do these  conclusions affect the validity of the calculations of Ginsparg and Perry?  They certainly do because the evidence is strongly that the semiclassical approximation they propose for the Euclidean path integral is not an approximation to the true behavior of the integral.  The reason is the dominance of infrared instabilities that invalidate assumptions about the behavior of the path integral required for the calculation to be well defined.  

\subsubsection{The sensitivity of Vilenkin's argument to the ultraviolet completion of general relativity}

Someone might object that the semiclassical calculations in question could not possibly be influenced by such technical issues, because they are insensitive to the ultraviolet completion of quantum gravity.  But it is easy to see that this cannot be the case, for the formula for black hole nucleation, (\ref{nucl1}) is dominated by Planck mass black holes.
So the ultraviolet completion of the theory is going to have a strong effect on the resulting amplitudes. (This is another example of uv-ir mixing.)  Whatever form of quantum gravity we have confidence in, we expect that there must be effects  which make it impossible to form black holes with masses greater than $M_{Pl}$.  But these effects are not seen in the semiclassical limit.  This implies that  quantum gravity effect, not captured by the naive semiclassical Euclidean path integral, which concern the existence and production of Planck scale black holes, are going to have a dominant effect on the actual black hole production rate, because the production of Planck scale black holes dominates exponentially in (\ref{nucl1}).     

One way to see this is give a much simpler and more physical derivation of the rate of nucleation of black holes in a thermal spectrum.   At any temperature 
$T<< T_{Pl}$
there will be a tail in the Planck distribution of quanta with arbitrarily high energy,
\f
\rho (E, T) = E^{-4} \frac{1}{e^{\frac{E}{T}}-1} \approx E^{-4} e^{-\frac{E}{T}}
\ff
 But for $E\approx E_{Pl}$ we expect, using conventional special relativity, that the
 associated wavelegth $\lambda = E^{-1}$ will be inside its Schwarzchild radius.
 $R_{Schw} = 2 G E$ and hence collapse to a black hole.  Hence, we expect
 from this hand waving argument a rate of black hole production in a thermal
 spectrum of order
 \f
 \Gamma \approx E^{-4}_{Pl} e^{-\frac{E_{Pl}}{T}}
 \ff
which agrees with (\ref{nucl1}).  

But this form of the argument is very sensitive to the form of the ultraviolet completion of the theory.  For example, there are proposals that quantum gravity effects modify the energy momentum relation so that there are no quanta of wavelength 
shorter than $l_{Planck}$\cite{DSR}.
These are the so called double or deformed special relativity.  We know in fact that such a modification holds in quantum gravity in $2+1$ dimensions coupled to matter\cite{DSR2+1} and there are reasons to expect it holds generally\cite{me-semi}.  In this case we would have the Planck spectrum cut off by $\lambda \approx l_{Pl}$.  The result could easily be a complete suppression of black hole production as that requires quanta with wavelengths $\lambda \leq l_{Pl}$.   

Note that if doubly special relativity is true, it is not an adjustable  feature that can be varied over a landscape. As shown in $2+1$ dimensions, it is forced by the ultraviolet consistency of the theory.  If the same holds in $3+1$ dimension than the result may be that no theory allows thermal production of black holes at finite positive $\Lambda$\footnote{To this must also be added the observation that the dominant population of black holes produced by the Ginsparg-Perry mechanism  evaporate almost instantaneously on the time scale, $t_N$ required to produce them.  This is very different from the astrophysical black holes involved in the original tests of CNS.  Whether this has an effect or not on the growth of the ensemble of universes is also a Planck scale problem, which should be investigated if Vilenkin's mechanism is to be further studied.}.

These considerations are not definitive but they are sufficient to show that Vilenkin's argument is highly speculative. It is just possible that all these objections might be evaded so that it hold in the true theory, but there are many reasons to suspect it does not. 
As a speculative argument it is of interest, but as a claim for ``falsification" of cosmological natural selection it has little force because it is so unlikely that the assumptions it makes are true at all, let alone apply to scales vastly bigger than the present Hubble scale. 

\subsubsection{Freak observers and the {\it reducto ad absurdum} of Vilenkin's argument.}

So far we have argued that it is not likely that Vilenkin's assumptions are realized in nature.
But there is a final issue to bring up, which is that if Vilenkin's argument works against cosmological natural selection, it works also against many predictions of Darwinian biology. 
Here is a minimal prediction of Darwinian  biology:

\begin{quotation}

{\bf D}  {\it Almost every $DNA$ sequence found in nature is found in the cells of a creature that is capable of reproducing itself. }

\end{quotation}

This is true in spite of the fact that almost every randomly chosen $DNA$ sequence does not code for any viable organism.  Indeed, since DNA does not exist except on planets with our kind of life, {\bf D} and many other predictions of Darwin's theory can be extended to the whole universe.   So Darwin's theory can be seen as a cosmological theory that makes many predictions about the distribution of objects  in the whole universe,  of the form of {\bf D}.  

Now, let us make the same assumptions as in Vilenkin's argument\cite{Alex},  including  that the universe evolves to the eternal deSitter vacua.  Due to its thermal flucutations there is a small, but non-vanishing  probability that a strand of DNA
coding any sequence $A$ comes into existence spontaneously.   We can estimate that the rate for this will have
the form
\f
\Gamma (A) \approx e^{-\alpha m(A)R}
\ff
where $m(A)$ is the mass of the DNA sequence and $\alpha >0 $ is a constant we do not need to estimate.  Since deSitter  spacetime is eternal after some time almost every DNA sequence in the history of the universe will be made by this spontaneous process. Since
$\Gamma (A)$ is insensitive to the viability or fitness of any organism
 {\bf D} is wrong.   By this schema one could argue that virtually any prediction of Darwin's theory is falsified  once one accepts the assumptions of Vilenkin's argument.  
 
Indeed, one could apply this reasoning to many predictions of science, because so long as there is a non-zero rate for a disconfirming instance to appear by a spontaneous thermal fluctuation, and so long as there are more disconfirming instances than confirming instances, which is the case for most predictions of science, the disconfirming instances will predominate in a universe that approaches the eternal deSitter vacua.   Therefor Vilenkin's argument could be used to discredit many  predictions of science which up till now were thought to be established.  

Of course we can easily get out of this problem by arguing that we are misapplying the principle of mediocrity. The predictions of Darwinian biology apply not to all DNA in the universe, but only to DNA sequences which are created in the cells of living creatures.
Certainly this is correct. But if this is allowed we can make exactly the same move to save the theory of cosmological natural selection. We can restrict attention to the ensemble of universes which are created from astrophysical black holes.  This evades Vilenkin's argument and returns CNS to the status it had before.  So, if we can save Darwinian biology by restricting the ensemble under consideration to that created by biological processes, we can just as easily save Darwinian cosmology by the analogous move.

It is remarkable that arguments of this kind have in fact appeared in the literature of the anthropic principle and eternal inflation\cite{freaks}.  This is because applying the principle of mediocrity requires making good on how to count the numbers of observers in a given universe.  But by an argument of the form just given, there is in eternal deSitter spacetime an infinite number of brains that appear spontaneously by a thermal fluctuation.  If they are counted then almost all observers arise as spontaneous fluctuations in the eternal deSitter vacua.  

This is a problem for the principle of mediocrity because we are then not typical observers.  Of course, one can then employ a version of the doomsday argument to predict that the universe must be unstable so as to avoid the conclusion.  But it is hard to see why the same argument, once allowed, would not hold against eternal inflation, in general.  

However,  if one discards the early collapse of the universe as highly improbable, but accepts the principle of mediocrity, then one can see the conclusion of the argument just given as a proof that the assumptions of Vilenkin's argument must be false.  That is, given that we are not ``freak brains",  the principle of mediocrity and the assumption that the universe evolves to an eternal deSitter vacuum cannot be both true. 

Of course one could always try to take advantage of the freedom offered by the principle of mediocrity to redefine the ensemble in which we are said to be ``typical".   But  if we are allowed to do this, we can use precisely the same  freedom  to redefine the ensemble considered in applications of Vilenkin's argument to CNS, as discussed above.    Thus, either Vilenkin's argument is wrong or it has no force against CNS.   

One might have thought that reasoning of this kind, which takes advantage of the flexibility built into the principle of mediocrity to choose the ensemble in which we are ``typical",  would be recognized as a {\it reducto ad absurdum} of a research program.  Instead there is in fact a literature about how the probability distribution used in application of the anthropic principle to eternal inflation can be modified to avoid the problem of the dominance of the class of observers by non-biological freaks\cite{freaks}.  
But by employing such a tactic, isn't one just demonstrating that  
cosmological scenarios such as eternal inflation that require some version of the principle of mediocrity to make our universe typical,  in a universe in which it is otherwise very untypical, are incapable of making unique predictions, because of the flexibility allowed by the ambiguity built into the notion of typicality?

\subsubsection{What the disagreement is about}

The last considerations suggest that there is a deeper aspect to this disagreement.  
In Darwinian biology we are not concerned with the far future, or what is eternally true, we are interested instead in explaining the past and the present and making predictions for
the immediate future.  Consequently we are concerned with relative fitness of nearby configurations, we do not try to discuss what a maximally fit organism would be like.  It is the same with the application of natural selection to cosmology.  Indeed, the application of natural selection to cosmology only makes sense in a context in which the universe is understood to be continually evolving to novel states at each time.   

By contrast, eternal inflation and the anthropic principle appear to be an attempt to restore
a picture of the universe which is static, in which we are concerned not with what is true
at any moment in our particular universe, but instead with properties of an eternal and static multiverse.  In this picture, things which are true are not bound in time, they refer instead
to populations defined over infinite, eternal time.   

This is a deep philosophical disagreement over how to approach the 
science of cosmology\footnote{For more about the philosophical and theological roots
of this disagreement, see \cite{LOTC}.}.  But what matters to science is only which is more successful and we will see over time which philosophy leads to genuine predictions and explanations.   My main claim is that it is the time bound, evolutionary picture, which is best suited to the world as we observe it, because what we observe is in fact a universe evolving continually in time.  By contrast, the main theoretical object one works with in eternal inflation and the antrhopic principle is not a representation of what we observe, it is an entirely invented eternal and static multiverse.  To go from the properties of this invented ensemble to what we observe in our universe one requires supplementary assumptions.
The principle of mediocrity is this supplementary assumption.  Its role, as we have seen above, is to pick out a distribution in which our universe is to be typical from an invented one in which it is not. As such it is required, but once it is admitted, the theory becomes too flexible, because there is an unavoidable freedom in choosing the precise ensemble within which our observed universe is to be typical.   

By contrast, CNS makes robust predictions in spite of our ignorance of details of the landscape, because it postulates a single mechanism that creates a single ensemble within which our universe must be typical.  This is why, at least so far, CNS has been more successful at making robust and falsifiable predictions for observations that are not only doable but in progress.  

All of this was anticipated by the American philosopher Charles Sanders Peirce, who wrote in  1891,

\begin{quotation}

{\it To suppose universal laws of nature capable of being apprehended by the mind and yet having no reason for their special forms, but standing inexplicable and irrational, is hardly a justifiable position.  Uniformities are precisely the sort of facts that need to be accounted for.  Law is par excellence the thing that wants a reason.  Now the only possible way of accounting for the laws of nature, and for uniformity in general, is to suppose them results of evolution\cite{CSP}.}

\end{quotation}

\section{Conclusions}

How are we to explain the choices of the parameters of the standard models of physics and cosmology?  How are we to account for the observation of special tuning to values that allow the existence of long lived stars and complex chemistry?  And can these be done in the context of a theory that makes genuine falsifiable predictions for doable experiments?

It seems that there is no evidence for the existence of a unique unified theory which gives unique predictions for the parameters.  Indeed, there is increased evidence for the existence of a landscape of string theories, as first postulated in \cite{strominger,evolve,LOTC}.  Nor does the anthropic principle offer much hope for a genuine explanation of the valued selected or for new falsifiable predictions.  

Instead, one can give conditions for a landscape based theory to succeed in generating falsifiable predictions for doable experiments as well as genuine explanations for the choices of parameters not based on anthropic reasoning.  These were reviewed in section 2 of this paper.  Cosmological natural selection was originally introduced in \cite{evolve} as an example of a cosmological scenario that satisfies these conditions.   To this day it appears to be the only proposed scenario of theory based on a landscape of theories that does succeed in making genuine falsifiable predictions for doable experiments.  We discussed above three such predictions, which remain verified by observations.  CNS is also the only genuine explanation for the choices of the parameters that solves the special tuning problem without resorting to anthropic reasoning.

After reviewing the status of CNS we discussed a very recent criticism of it proposed by Vilenkin.  We first strengthened Vilenkin's argument, so as to make it not dependent on reasoning about infinite quantities.  Then we showed why the reasoning involved is unlikely to be reliable.  The reasons included,

\begin{itemize}

\item{} The Vilenkin argument requires that there be no novel physical phenomena other than  a simple cosmological constant when physics is extrapolated sixty orders of magnitude beyond presently observable scales.  But there are already indications in CMB data that new physics is required at the present Hubble scale. 

\item{}The Vilenkin argument only works in a single one of several modifications of general relativity motivated by the observation of accelerated expansion.   It would not work in the others.

\item{} The Vilenkin argument relies on calculations using semiclassical methods in Euclidean path integrals, vintage early 1980's.   These calculations appear to be semiclassical but they are in fact exponentially sensitive to new physics at the Planck scale because the production rates which result are dominated by Planck scale processes.  The results are then very vulnerable to modifications at Planck scales including cutoffs and modifications of energy momentum relations at Planck scales that would greatly suppress or forbid the production of Planck scale black holes.

\item{}The use of the Euclidean path integral made by Vilenkin assumes that thermal equilibrium has been established among the gravitational degrees of freedom.  But there are compelling arguments that thermal equilibrium is never reached for gravitons.  

\item{}  The theoretical basis  for believing in the reliability of such semiclassical Euclidean calculations is also very much weakened by results that show that the Euclidean path integral in quantum gravity defines theories in different universality classes than the Lorentzian path integrals.  

\item{} The Vilenkin argument could be used against many successful predictions of Darwinian biology. To avoid this one takes advantage of the flexibility built into the principle of mediocrity to restrict attention to DNA which is the result of biological evolution. But the same move also allows us to restrict attention to the ensemble of universes created in astrophysical black holes, thus making CNS safe from Vilenkin's critique as well as from the criticism that the multiverse may be dominated by universes created in eternal inflation.

\end{itemize}

It is always good to challenge ideas with novel criticisms, and Alex Vilenkin is to thanked for presenting a challenge to cosmological natural selection.   But a careful look shows that his challenge is based on assumptions that are not necessarily reliable, hence the challenge is not very strong.  CNS remains vulnerable to falsification, and so not only is a viable cosmological model, it continues to illustrate the necessity of the conditions stated in section 2.1 for making falsifiable predictions from a theory based on a landscape.    

\section*{Acknowledgements}

I am grateful to  Hilary Greaves, Andrei Linde, Laura Mersini and Steve Weinstein for conversations about these issues and to Sabine Hossenfelder and Alex Vilenkin for comments on the  manuscript.  
Research at PI is supported in part by the Government of 
Canada through NSERC and by the Province of Ontario 
through MEDT.

\end{document}